# Test Results of PIP2IT MEBT Vacuum Protection System*

A. Chen†, R. Andrews, C. Baffes, D. Lambert, L. Prost, A. Shemyakin, T. Zuchnik

Fermilab, Batavia, IL 60510, USA

*Abstract*

The central part of PIP-II program of upgrades proposed for the Fermilab injection complex is an 800 MeV, 2 mA, CW-compatible SRF linac. Acceleration in superconducting cavities begins from a low energy of 2.1 MeV, so that the first cryomodule, Half Wave Resonator (HWR) borders the warm Medium Beam Transport (MEBT) line. To minimize the amount of gas that may enter the SRF linac in a case if a vacuum failure occurs in the warm front end, a vacuum protection system is envisioned to be used in the PIP-II MEBT. It features a fast closing valve with two sensors and a differential pumping insert. The system prototype was installed in the PIP-II Injector Test (PIP2IT) accelerator and successfully tested in several modes modelling the vacuum failures. The report presents the design of the vacuum protection system and results of its tests.

## INTRODUCTION

As the injection test experiment of PIP2, PIP2IT is composed of an H- ion source (IS), low energy beam transfer (LEBT), RFQ, medium energy beam transfer (MEBT), a cryomodule of half wavelength resonator (HWR), a cryomodule of single spoke cavity resonator (SSR1), and high energy beam transfer (HEBT) with beam dumper. HWR and SSR1 are cryomodules with superconducting RF cavities operating under 2K. furthermore the performance quality of superconducting cavity critically relies on the quality of low particulate and ultra-high vacuum environment. Since the beam line vacuum space is in common for all beam devices, the inevitable vacuum failure in warm section pose significant risk on SRF cavities. In case of serious vacuum failure, the large gas flux into cryomodule not only destroy the superconducting status, but also moving the particles from adjacent area into superconducting cavity and ruin its performance. It is necessary to equip the warm section adjacent to cryomodules, such as MEBT, HEBT with measure to isolate the gas propagation in vacuum at the speed of hundreds meter per second.

## THE DESIGN OF VACUUM PROTECT SYSTEM

A typical way to achieve this isolation is to utilize the fast closing valve, such as the one by VAT. The spec of VAT's 75 series fast close valve can close with 10ms. At PIP2IT, in order to prevent large amount of gas (and particles will move with) flux into HWR (the 1$^{st}$ cryomodule of PIP2) during any possible vacuum failure in MEBT, the fast valve is placed about 1 m upper stream of HWR, the 1$^{st}$ sensor that detecting very large vacuum pressure rising (it was set to 1E-4 torr) will be placed at the beam absorber. And the 2$^{nd}$ sensor will be placed at the bunch cavity near HWR. The absorber will dump all chopped beam where large amount of gas is produced, and particle is created. A differential pumping insert (DPI) with small aperture of 10mm in dia., 200mm long is placed downstream of absorber to significantly reduce the gas flux from any possible vacuum failure in upstream. The DPI helps to achieve UHV in the region next to HWR as well. the pressure rose detected from either sensor will trig the fast valve close within 10ms. Since gas propagation in vacuum is very fast, closing valve within 10ms is not fast enough to completely prevent gas flux pass the valve before it close, especially if a failure occurs near the valve. The test for measuring this amount of gas past the valve was carried out to qualify if the design can provide meaningful protection to HWR.

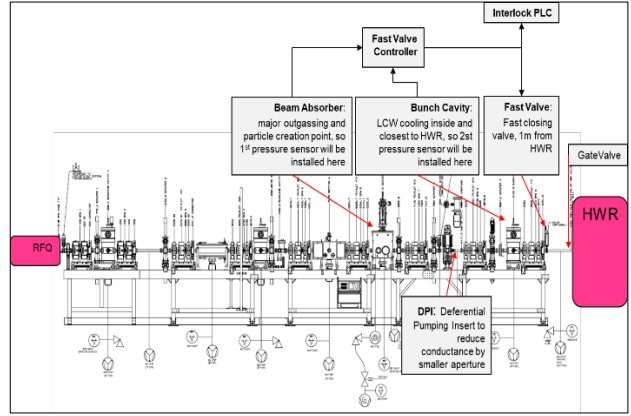

Figure 1: The configuration of Vacuum Protection System at MEBT

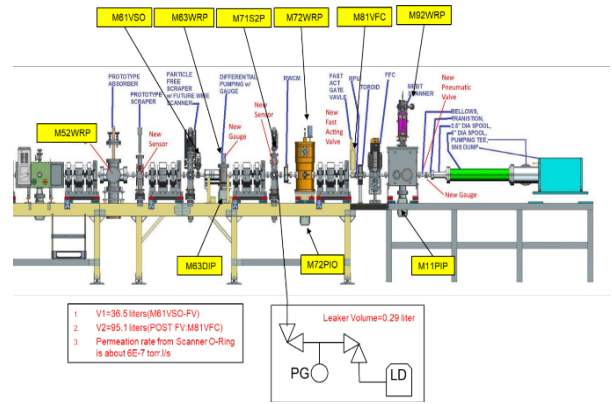

Figure 2: Setup of 1$^{st}$ Test

however, the closing time was not measured.

## 1ST TEST SETUP AND RESULTS

Along with the progress of PIP2IT, the 1st measurement was carried out in less ideal configuration. The 1st sensor was installed on the scraper near protype absorber, and the 2nd was the scraper near the bunch cavity. The DPI was placed in between. In this setup, the vacuum space at upper stream of the fast valve is 36.5 litters, and 95.1 litters at the downstream. The large volume at the downstream, where the gas amount measured, reduced the sensitivity of the test, furthermore, the permeation from a large O-Ring in the emittance scanner set the low limit of gas flux rate to E-9 torr.l/s.

The leak was simulated by a fixed volume gas reservoir

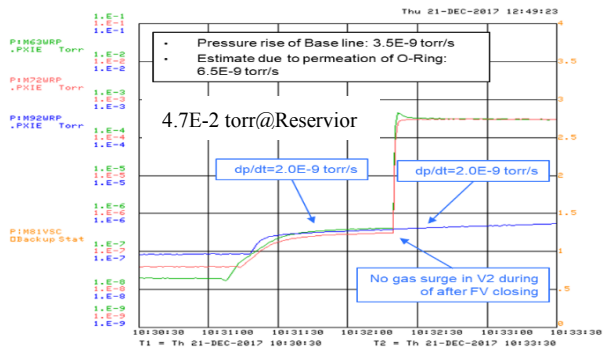

Figure 3: Sample result of 1st Test,

(0.29 litter in 1st test) and designated pressure. The leak was introduced by opening the manual valve into vacuum space in upper stream of the fast valve. The leak was placed at one of the two scrapers where the sensor is, but from opposite ports.

From the 1st test, we learned the gas amount past the fast closing valve is smaller than expected, below the permeation rate from O-Ring.

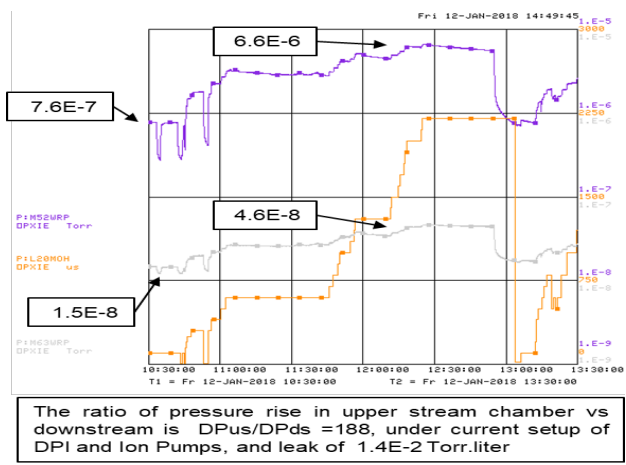

Figure 4: Effect of DPI from 1st Test

We also learned from the 1st test that the DPI efficiently restrict the gas flux. The effectiveness of DPI decreased as the leak size increase.

In order to improve the sensitivity of measurement, it is necessary to reduce the vacuum space at the downstream of the fast valve, even more important is to eliminate the device with O-Ring.

## 2ND TEST AND RESULTS

In the 2nd test setup, the vacuum space was reduced to 2.79 litters, as in Fig. 5. There is no O-ring this vacuum space so no air permeation. The DPI effect of flux restriction is similar as in tests in 1st setup. The ion pump

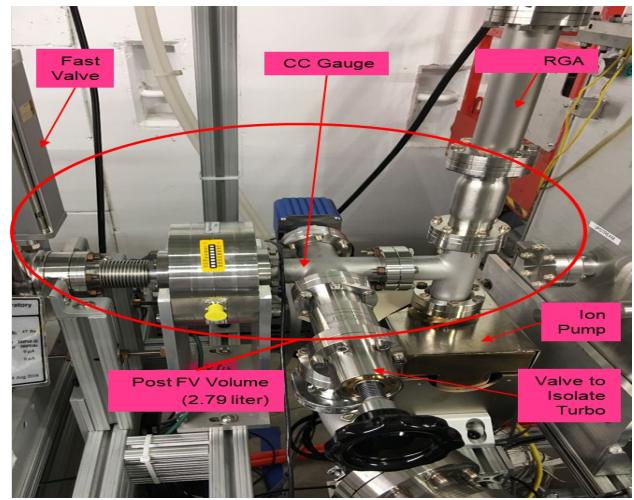

Figure 5: Vacuum space at downstream of Fast Valve

was off during leak test, it was used only for recovering vacuum.

Fig. 6 shows the setup of simulated leak. The dry nitrogen gas reservoir has volume of 0.33 litter. Leak size changes by varying the pressure from 1.2 torr to 760 torr, and 760 torr with continuous supply.

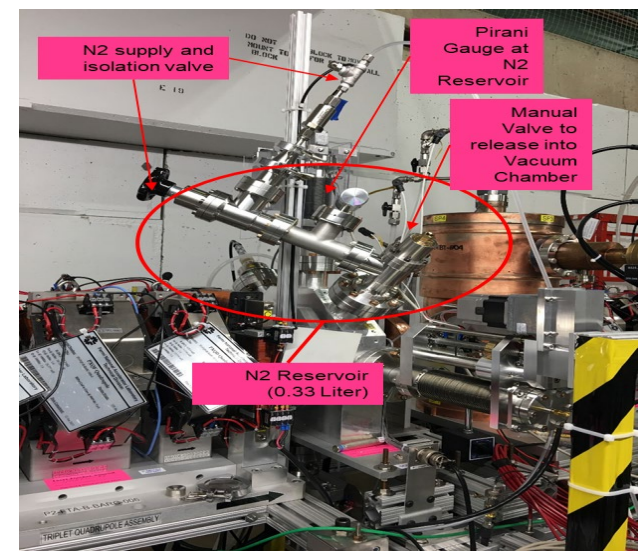

Figure 6: Simulated Leak setup

Fig.7 shown a typical response from a simulated leak. The pressure at leak indicates the size of leak since the volume of reservoir is fixed. Pressure rise response the leak promptly.

There are ten measurement took in the 2$^{nd}$ test. Various

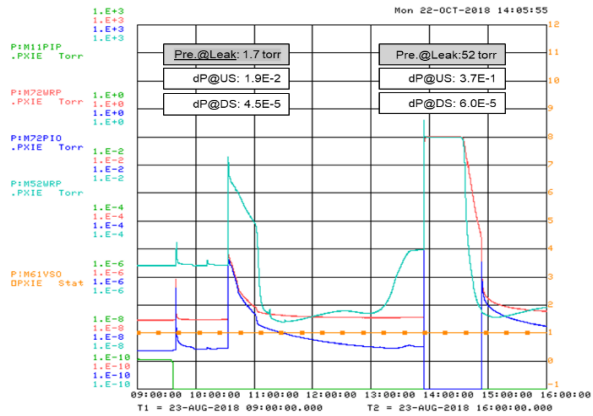

Figure 7: A sample of measurement at 2nd test

leak sizes were simulated in both upper stream and downstream of the DPI, by changing the pressure of dry nitrogen reservoir. The gas pressure in the vacuum space downstream of fast valve was reading from the cold cathode gauge for every simulated leaking. P0 was the base pressure before leak, P1 was the peak pressure right after simulated leak induced. Pressure rise was P1 minus P0. The gas amount past the fast valve then equals the pressure rise times the fixed volume. Monolayer coverage

Table : Summary of measurement for 2$^{nd}$ test

| Leaker Reservoir | CCG500 Reading P0 (before) | P1 (after) | Pressure Rise | Gas Amount | monolayer coverage | Leaker Location |
|---|---|---|---|---|---|---|
| torr | torr | torr | torr | torr.liter | cm$^2$ | |
| 23-Aug 1.7 | 6.2E-09 | 1.7E-07 | 1.6E-07 | 4.6E-07 | 1.3E-02 | US DPI |
| 52 | 7.5E-09 | 1.9E-07 | 1.8E-07 | 5.1E-07 | 1.5E-02 | |
| 760 | 7.9E-09 | 2.0E-07 | 1.9E-07 | 5.4E-07 | 1.6E-02 | |
| 24-Aug 760 | 2.3E-08 | 2.3E-07 | 2.1E-07 | 5.8E-07 | 1.7E-02 | |
| 1-Oct 9.5 | 7.3E-09 | 3.8E-05 | 3.8E-05 | 1.1E-04 | 3.1E+00 | DS DPI |
| 350 | 2.1E-08 | 1.4E-05 | 1.4E-05 | 3.9E-05 | 1.1E+00 | |
| 810 | 1.0E-07 | 4.5E-05 | 4.5E-05 | 1.3E-04 | 3.6E+00 | |
| 8-Oct 1.2 | 6.1E-09 | 5.1E-04 | 5.1E-04 | 1.4E-03 | 4.1E+01 | |
| 9-Oct 130 | 6.0E-09 | 5.8E-07 | 5.7E-07 | 1.6E-06 | 4.6E-02 | |
| 760 | 3.0E-08 | 2.5E-06 | 2.5E-06 | 6.9E-06 | 2.0E-01 | |

is calculated by the amount of gas covers the surface with one molecular layer.

From the measurement for ten results, we observed that 1) the size of leaks is insensitive to the amount of gas past through the fast valve before close, likely due to all simulated leak size too large; 2) the DPI plays significant role to reduce the amount of gas can pass the fast valve before it close, two groups of leak locations, upstream and downstream of DPI is clearly separated; 3) in term of monolayer coverage, the gas amount is reasonable small to cavity surface area.

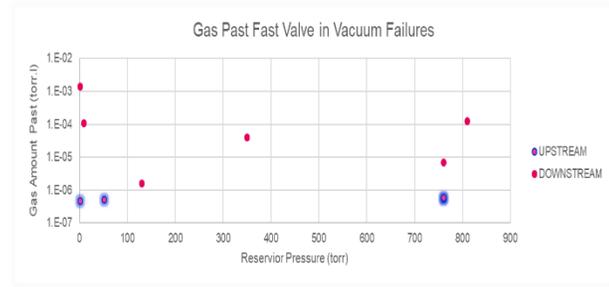

Figure 8: Gas amount vs Leak Size

## SUMMARY

With the improved sensitivity in 2$^{nd}$ test setup, the amount of gas past the FV is carefully measured. It is not directly driven by the size of leak. Differential Pumping Insert (DPI) throttled the air leak significantly, however its effect decreased as leak size increased. The current design works as expected.

The amount of gas past FV is small enough 1) not able to move particulates, 2) insignificant for surface condensation of cavities. So, the vacuum protection system can provide significant protection to cryomodule from the risk of vacuum failure in warm sections.

## ACKNOWLEDGMENT

The authors are thankful to the many people who built the PIP2IT MEBT and helped with these tests, including R. Kellett, E. Lopez, C. Exline.